\documentclass[%
 reprint,
superscriptaddress,
groupedaddress,
 amsmath,amssymb,
 aps,
pra,
prb,
rmp,
prstab,
prstper,
floatfix,
]{revtex4-1}

\usepackage{graphicx}% Include figure files
\usepackage{dcolumn}% Align table columns on decimal point
\usepackage{bm}% bold math
\usepackage{hyperref}% add hypertext capabilities
\usepackage{placeins}
\usepackage{xcolor}
\usepackage{ulem}

\begin{document}

\preprint{APS/123-QED}

\title{Embedded Topological Semimetals}

\author{Saavanth Velury}
\author{Taylor L. Hughes}
\affiliation{Department of Physics and Institute of Condensed Matter Theory, University of Illinois at Urbana-Champaign, IL 61801, USA}

\date{\today}% It is always \today, today,
             %  but any date may be explicitly specified

\begin{abstract}
Topological semimetals, such as Dirac, Weyl, or line-node semimetals, are gapless states of matter characterized by their nodal band structures and surface states. In this work, we consider layered (topologically trivial) insulating systems in $D$ dimensions that are composed of coupled multi-layers of $d$-dimensional topological semimetals. Despite being nominal bulk insulators, we show that crystal defects having co-dimension $(D-d)$ can harbor robust lower dimensional topological semimetals embedded in a trivial insulating background. As an example we show that defect-bound topological semimetals can be localized on stacking faults and partial dislocations. Finally, we propose how an embedded topological Dirac semimetal can be identified in experiment by introducing a magnetic field and resolving the relativistic massless Dirac Landau level spectrum at low energies in an otherwise gapped system.
\end{abstract}

\pacs{Valid PACS appear here}% PACS, the Physics and Astronomy
                             % Classification Scheme.+
                             
%\keywords{Suggested keywords}%Use showkeys class option if keyword
                              %display desired
\maketitle

%\tableofcontents

\section{I\lowercase{ntroduction}}
\indent Over the past decade, the classification of gapped, non-interacting topological phases of matter has been extensively developed. Starting with strong topological insulators protected by only time-reversal, particle-hole, and/or chiral symmetries\cite{Kane2005,Kane2005a,Bernevig2006,Schnyder2008,Kitaev2009,Qi2009,Ryu2010,hasan2010colloquium,Qi2011}, the classification has grown to include weak topological insulators (those protected by translation symmetries)\cite{Fu2007,Moore2007,Roy2009} as well as those protected by crystalline symmetries\cite{Fu2011,fang2012bulk,Chiu2013,fang2013entanglement,Jadaun2013,Slager2013,Benalcazar2014,Shiozaki2014,kruthoff2017topological,song2017topological,po2017symmetry,bradlyn2017topological,cano2018building,po2018fragile,song2018quantitative,Ono2020,po2020symmetry,song2020twisted,Elcoro2020,Ono2020a}.Mirroring this development in topological insulators, the classification of topological semimetals has seen recent rapid development\cite{Yang2014,Bradlyn2016,Watanabe2016,Wu2021} building upon earlier work that explored their exotic transport properties\cite{Hosur2012,Hosur2013,Wang2017}, robust anomalous surface states\cite{Wan2011,Matsuura2013,Xu2015,Bian2016a,Kargarian2016,Wieder2020,Obakpolor2021}, and electromagnetic responses\cite{Shi2007,Zyuzin2012,Vazifeh2013,Haldane2014,ramamurthy2015patterns,Ramamurthy2017,Armitage2018,Robredo2021,Gioia2021,Pal2021}. 

Unlike their insulating counterparts, topological semimetals are gapless in nature, and their topology is characterized by their symmetry/topology protected band degeneracies (nodes) near the Fermi level. In general, topological semimetals can be classified as Dirac, Weyl, or line-node based on the dispersion of their nodes, and what symmetries are required to protect the degeneracies. This classification has become more refined through consideration of other features such as the number of degeneracies, the resulting topology of the Fermi surface, their distribution throughout the Brillouin zone (BZ), etc. Additionally, the progress made in the theoretical examination of topological semimetals has been aided by their experimental discovery in real materials. Some notable examples include the three-dimensional Dirac semimetals Na$_{3}$Bi\cite{Liu2014} and Cd$_{3}$As$_{2}$\cite{Liu2014a,Neupane2014}, the Weyl semimetal TaAs\cite{Xu2015}, and line-node semimetals PbTaSe$_{2}$\cite{Bian2016} and TlTaSe$_{2}$\cite{Bian2016a}. There have also been many material predictions made for topological semimetals based on first-principles calculations\cite{Xu2011,Burkov2011a,Wang2012,Young2012,Steinberg2014,Bulmash2014,Chen2017} that will likely lead to new experimental discoveries.

\indent The connection between symmetry-protected topology and defect sensitivity is fundamental to the study of topological insulators\cite{teo2017topological}. Previous work has unveiled a rich set of phenomena associated to the generation of stable fermionic bound states localized on various types of topological defects, e.g., vortices, dislocations, or disclinations\cite{Ran2009,Teo2010,teo2013existence,Benalcazar2014,Queiroz2019,Li2020,Roy2020,Nag2020}. Interestingly, a recent work\cite{tuegel2018embedded} provided evidence of a nominally topologically trivial insulator capable of hosting topological features on crystal defects if it is formed from lower-dimensional layers of topologically non-trivial insulators. While the total topology of the entire layered system is trivial, the topology of isolated layers can be revealed with stacking faults or partial dislocations, which hence give rise to protected surface states at the boundaries of the defect. This concept was dubbed an embedded topological insulator (ETI) - topological insulators embedded as defects in (topologically) trivial insulating environments of greater dimensionality. 

\indent It was subsequently shown in later works\cite{Khalaf2018a,Matsugatani2018,Trifunovic2019} that higher-order topological insulators (HOTIs) can be smoothly deformed to ETIs (and vice versa) via a procedure that preserves the bulk gap and global crystalline symmetry at the expense of breaking translational and local crystalline symmetries. Given a $d$-dimensional HOTI hosting an $n^{\text{th}}$-order gapless boundary mode ($d>n$), an ETI can be generated by trivializing the regions of the HOTI adjacent to this boundary mode, resulting in a gapped topological phase with codimension $n-1$ embedded in a trivial bulk, which is an ETI. A specific example of this is considered in\cite{Matsugatani2018} for a 3D inversion-symmetric HOTI composed of a coupled layers of Chern insulators with alternating Chern number $C=\pm 1$, with a single Chern insulator of $C=+1$ containing the inversion center. The Chern layer containing the inversion center hosts the chiral edge mode that corresponds to the hinge mode of this HOTI. One can smoothly adjust the inter-layer coupling to dimerize the layers above and below the inversion center, resulting in gapless edge modes with opposite chirality becoming gapped out and effectively trivializing the bulk halves neighboring the inversion center. The residual Chern insulator containing the inversion center is precisely a 2D embedded Chern insulator.

In light of these results, it is natural to ask if topological semimetals can emerge in a trivial, bulk insulating environment through the introduction of defects, i.e., can we have robust, embedded topological semimetals? A precursor to this question was answered in the context of topological band insulators (TBIs) in Ref.~\onlinecite{Slager2016}, which demonstrated that grain boundaries in TBIs can host two-dimensional time-reversal symmetry protected semimetals. From this, it is natural to ask if other types of defects, such as stacking faults or dislocations, can give rise to novel embedded topological (semi)metallic phases that would be unexpected in a bulk insulator. Indeed, since transport is a global measurement, such defect-generated conductivity would show up as seemingly spurious transport signals in otherwise insulating system.

In this work, we answer these questions by extending the idea presented in Ref.~\onlinecite{tuegel2018embedded}. Namely, we introduce the notion of an embedded topological semimetal (ETSM). The main focus of our work is on layered (topologically trivial) insulating systems in $D$-dimensions that are composed of coupled multilayers of $d$-dimensional topological semimetals. We will consider the effects of two types crystalline defects: stacking faults and partial dislocations. We demonstrate that when these defects are introduced in these nominal bulk insulators, defect-bound lower dimensional topological semimetals can emerge, embedded in a trivial insulating background. Furthermore, we show that when these defects host 2D topological Dirac semimetals their presence can be detected by observing the characteristic Landau level spectrum for massless Dirac fermions. 

This article is organized as follows. In Sec.~\ref{sec:embeddedmetal}, we initially build up the concept of an ETSM by first presenting a model for embedded metals occurring in a $D=2$-dimensional (2D) insulating system composed of layered $d=1$-dimensional (1D) spinless electron chains coupled to each other. By introducing two stacking faults at a set distance apart from each other, the system transitions from being a bulk insulator having metallic behavior near the defects. Then, in Sec.~\ref{sec:embeddeddiracsemimetal}, we introduce our definitive example of an ETSM: an embedded topological Dirac semimetal. The construction involves a $D=3$-dimensional (3D) topologically trivial insulating system composed of layered 2D topological Dirac semimetals. We demonstrate that introducing a stacking fault or partial dislocation in this system results in a topological Dirac semimetal in an otherwise trivially gapped system. In Sec.~\ref{sec:landaulevel}, we discuss how an ETSM can be experimentally determined by introducing a magnetic field and resolving the relativistic massless Dirac Landau level spectrum at low energies in an otherwise gapped system. Finally, in Sec.~\ref{sec:conclusion}, we offer concluding remarks. 
\section{E\lowercase{mbedded} M\lowercase{etals} \lowercase{in} (1+1)-D}
\label{sec:embeddedmetal}
\begin{figure}[ht]
\includegraphics[scale=0.325]{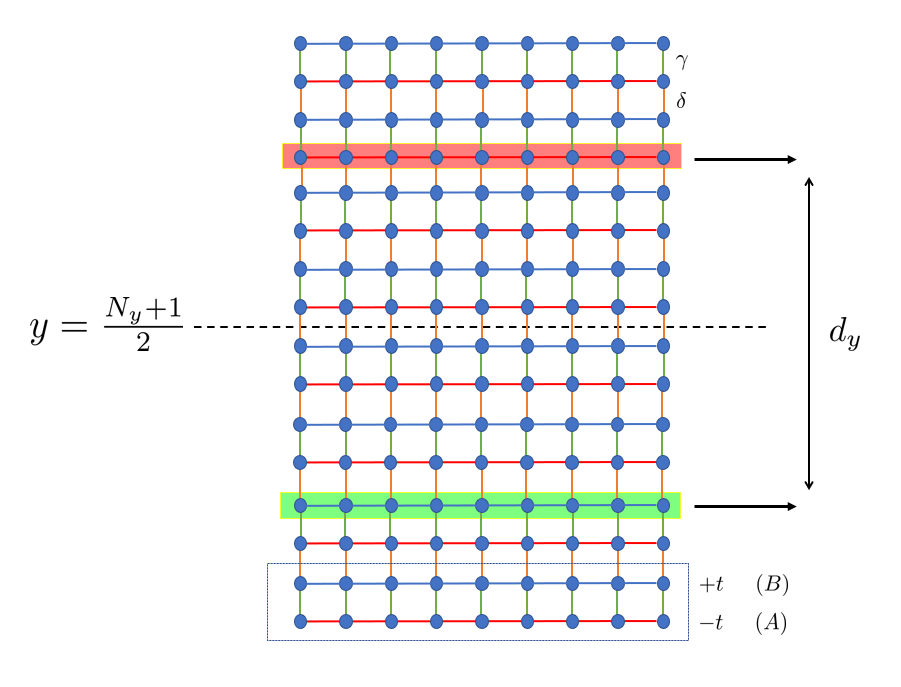}
\caption{Illustration of the stacking fault procedure. The pristine bulk insulator described by the Bloch Hamiltonian given by Eq.~(\ref{eq:pristineBlochhamiltonian2D}) is shown above for $N_{y}=8$ unit cells (indicated by the dashed outlines). Each unit cell contains two spinless 1D electron chains with hopping $-t$ (denoted by the horizontal red colored bonds, labeled $A$ within each unit cell) and $+t$ (denoted by the horizontal blue colored bonds, labeled $B$ within each unit cell). The intra-cell coupling strength is denoted by $\gamma$ (vertical green colored bonds in each unit cell) and the inter-cell coupling strength is denoted by $\delta$ (vertical orange colored bonds between unit cells). Two chains are extracted, one of type $A$ and the other of type $B$ as highlighted.}
\label{fig:embeddedmetal}
\end{figure}

\indent We begin by introducing the model of embedded metals in a 2D insulating system. This section serves as a warmup to the concept of an ETSM. Although the embedded metals that we explain in detail in this section are topologically trivial and therefore, not ETSMs, it serves as an example of the importance of translational-symmetry breaking required to realize the embedded nature of these systems that we will discuss throughout this work. Specifically, we consider stacking faults introduced at certain locations on the real-space lattice and directly correspond them to the specific bands in the bulk band structure that intersect the Fermi level. We will show that these metallic bands arise directly from the exposed lower dimensional metallic layers where the stacking faults are introduced, which we identify as the embedded metals.\\
\indent First, we will describe the pristine model when no stacking faults are present. This model consists of 1D spinless electron chains extended in the $x$-direction, and stacked and coupled to each other along the $y$-direction. Each one-dimensional chain is described by the Bloch Hamiltonian
\begin{equation}\begin{split}\label{eq:spinlessblochhamiltonian}
h_{\text{1D}}(k_{x})=\pm 2t\cos(k_{x}),
\end{split}\end{equation}
where $t$ is the strength of the nearest-neighbor hopping, and we have set the lattice constant $a=1$. Layering these chains on top of each other, we place two chains with opposite hoppings $+t$ and $-t$ within each unit cell along the $y$-direction, with a total of $N_{y}$ unit cells in the $y$-direction. This is illustrated in Fig.~\ref{fig:embeddedmetal}. We denote the chain with hopping $-t$ with the label $A$ and the other with hopping $+t$ with the label $B$.
\begin{figure}[ht]
\includegraphics[scale=0.35]{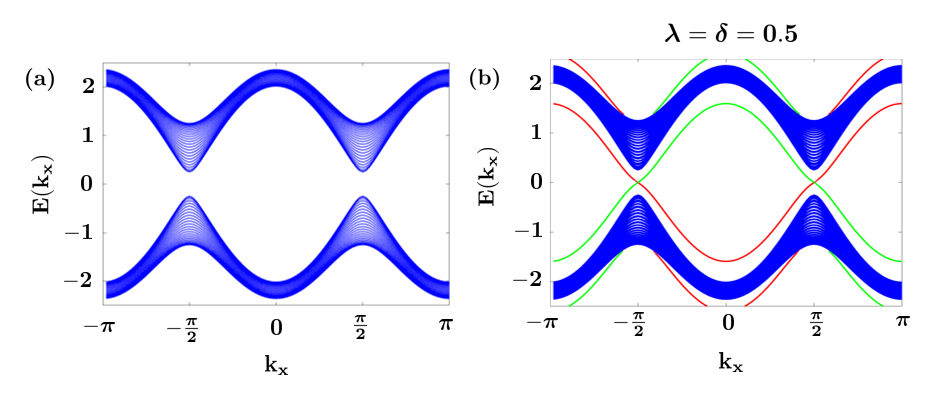}
\caption{Band structure of the pristine bulk insulator given by Eq.~(\ref{eq:pristineBlochhamiltonian2D}) shown in (a), for $\gamma=0.75$, $\delta=0.5$, and $t=1$ for $N_{y}=50$ unit cells ($100$ spinless electron chains), with periodic boundaries in the $y$ direction. In (b), the band structure of an embedded metal after the stacking faults are introduced at $n_{y}=6^{\text{th}}$ (a $B$ chain is extracted, the effect on the band structure is highlighted by the green bands) and at $N_{y}+1-n_{y}=45^{\text{th}}$ unit cells (an $A$ chain is extracted, the effect on the band structure is highlighted by the red bands). The exposed chains are coupled with strength $\lambda=\delta=0.5$.}
\label{fig:embeddedmetalenergyspectrum}
\end{figure}

The Bloch Hamiltonian describing this model is given by $H(\mathbf{k})=\vec{\mathbf{d}}(\mathbf{k})\cdot\vec{\boldsymbol{\sigma}}$ with
\begin{equation}\begin{split}\label{eq:pristineBlochhamiltonian2D}
\vec{\mathbf{d}}=(\gamma+2\delta\cos(k_{y}),2\delta\sin(k_{y}),-2t\cos(k_{x})),
\end{split}\end{equation}
where $\gamma$ is the intra-cell hopping between the chains, $\delta$ is the inter-cell hopping, and $\vec{\boldsymbol{\sigma}}=(\sigma_{x},\sigma_{y},\sigma_{z})$ are the Pauli matrices expressed in the wire basis within each unit cell (i.e., $A$ and $B$). The pristine Hamiltonian has a spinless time-reversal symmetry given by $T=K$ where $K$ is the complex conjugation operator ($TH(\mathbf{k})T^{-1}=H^*(\mathbf{k})=H(-\mathbf{k})$). The energy spectrum of (\ref{eq:pristineBlochhamiltonian2D}) is given by $E(\mathbf{k})=\pm|\vec{\mathbf{d}}|=\pm\sqrt{(\gamma+\delta\cos(k_{y}))^{2}+(\delta\sin(k_{y}))^{2}+(-2t\cos(k_{x}))^{2}}$. Thus, for nonzero $\gamma$ and $\delta$, the spectrum of the Hamiltonian is gapped for all $|\gamma|\neq|\delta|$, and therefore a bulk insulator. The band structure of the pristine bulk insulator is illustrated in Fig.~\ref{fig:embeddedmetalenergyspectrum} (a).

 Now we will consider the effects of introducing stacking faults, as illustrated in Fig.~\ref{fig:embeddedmetal}. We generate stacking faults by removing a spinless electron chain of type $A$ and another of type $B$ at two separate locations, which will introduce two inequivalent stacking faults. A chain of type $B$ is removed from the $n_{y}^{\text{th}}$ unit cell where $1\leq n_{y}<N_{y}/2$, and a chain of type $A$ is removed from the $(N_{y}+1-n_{y})^{\text{th}}$ unit cell. The two chains that neighbor each of the removed lines are then coupled with strength $\lambda$. The resulting energy spectrum after this procedure is illustrated in Fig.~\ref{fig:embeddedmetalenergyspectrum} (b). We find two metallic dispersing bands that lie in the insulating gap, one arising from, and localized at, each stacking fault. There are gapless Fermi points ($E_{F}=0$) at $k_{x}=\pm\frac{\pi}{2}$. Intuitively, this occurs because the exposed $A$ and $B$ chains in the $n_{y}^{\text{th}}$ and $(N_{y}+1-n_{y})^{\text{th}}$ unit cells respectively each have energy dispersion given by $E_{\text{1D}}^{(\pm)}(k_{x})=\pm 2t\cos(k_{x})$, with the exposed $A$ chain carrying the positive dispersion and the exposed $B$ chain carrying the negative one. The metallic dispersions persist for any value of $\lambda\in[0,\max(\gamma,\delta)]$, not just $\lambda=\delta$. However, the Fermi velocity of the band dispersion around $k_{x}=\pm\frac{\pi}{2}$, decreases as $\lambda$ is increased. Additional plots demonstrating this, and a $k\cdot p$ expansion of the dispersion around $k_{x}=\pm\frac{\pi}{2},$ are provided in Appendix A. Therefore, the exposed $A$ and $B$ chains lying on the stacking faults are denoted as the embedded metals of the system. Thus, this example is a clear illustration that a metallic system can be realized from a bulk insulator through stacking fault defects that expose the underlying embedded metals within the system.
 
\section{E\lowercase{mbedded} T\lowercase{opological} D\lowercase{irac} S\lowercase{emimetal} \lowercase{in} (2+1)-D}
\label{sec:embeddeddiracsemimetal}
\indent Having established how a bulk insulator can be turned into a metallic system by introducing stacking faults, we now turn our attention $D=3$ dimensions. Our paradigmatic example is the embedded topological Dirac semimetal (ETDSM), which we will establish in this section. We will consider a 3D nominal bulk insulator constructed from 2D topological Dirac semimetal layers. First, we will establish the technical details of the model, including the layer Hamiltonian and its phase diagram, and then the full model Hamiltonian built from coupling the individual layers. Then, we establish how the stacking fault is introduced into our system, and illustrate the change in the bulk band structure, which transforms from being gapped to becoming gapless with robust Dirac nodes. We identify these Dirac nodes as those arising from the exposed 2D topological Dirac semimetal layer lying on the stacking fault, which is the ETDSM. Finally, we will characterize the topology of the resulting ETSM by analyzing the Berry phase characteristics across the BZ. In this discussion, we note that the symmetries required for topological protection of the Dirac nodes of the ETDSM are the same symmetries required to protect the Dirac nodes of the individual topological Dirac semimetal layers.
\begin{figure}[ht]
\includegraphics[scale=0.25]{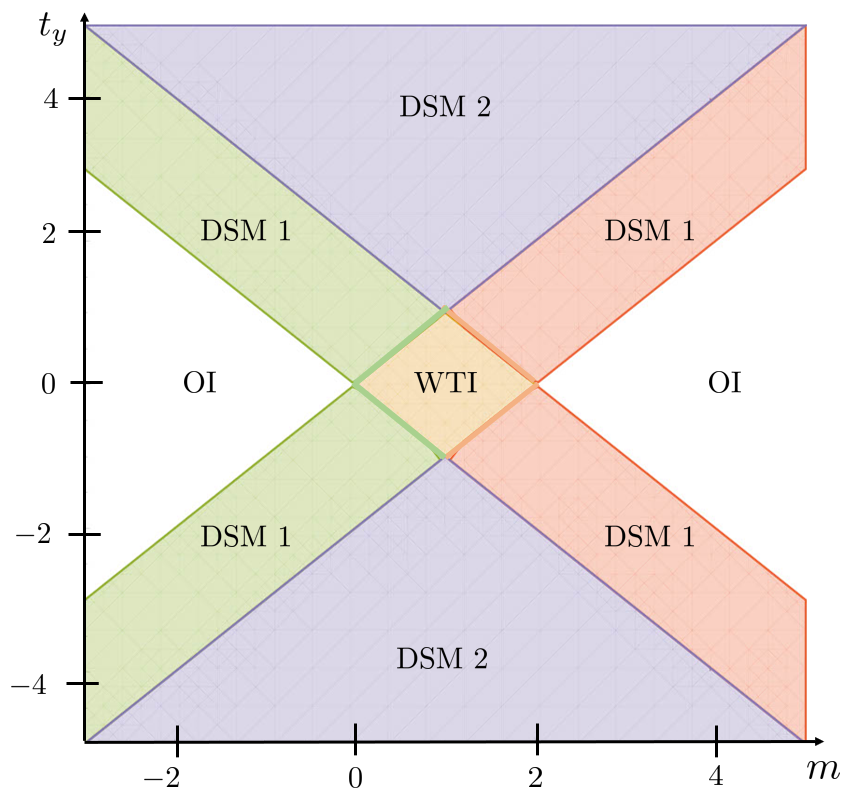}
\caption{Phase diagram of the 2D layer Hamiltonian given by Eq.~(\ref{eq:2DWTI}) for model parameters $m$ and $t_{y}$. The phases of this model include the ordinary insulator (OI) phase, weak topological insulator (WTI) phase, Dirac semimetal 1 (DSM 1) phase, and Dirac semimetal 2 phase (DSM 2) phase. The DSM 1 phase contains one pair of Dirac nodes as opposed to two pairs of Dirac nodes in the DSM 2 phase.}
\label{fig:2DWTIphasediagram}
\end{figure}
\subsection{Layer Hamiltonian and Phase Diagram}
\label{subsec:layerhamiltonian}

Each of the 2D layers comprising our full 3D model are adapted from a square lattice Dirac semimetal model from Ref.~\onlinecite{ramamurthy2015patterns}, the details of which we briefly review in this subsection. The 2D Dirac semimetal layer itself is constructed from coupling 1D topological insulator chains having Bloch Hamiltonian:
\begin{equation}\begin{split}\label{eq:1DTI}
H_{\text{1DTI}}(k_{x})=\sin(k_{x})\sigma_{y}+(1-m-\cos(k_{x}))\sigma_{z},
\end{split}\end{equation}
where $m$ is a parameter controlling the insulating phases of the model, and the Pauli matrices act on spinless orbitals labeled by $1$ and $2$ respectively. This model has an inversion symmetry given by $I=\sigma_{z}$ ($IH_{\text{1DTI}}(k_{x})I^{-1}=H_{\text{1DTI}}(-k_{x})$), which can protect an inversion symmetric topological insulator phase\cite{hughes2011inversion}. The important topological invariant that characterizes this model is $\mathbb{Z}_{2}$ in nature, and is captured by the bulk charge polarization $P$. For $m<0$ or $m>2$, the bulk charge polarization is given by $P=0\hspace{0.1cm}(\text{mod}\hspace{0.05cm}e),$ and for $0<m<2$, $P=\frac{e}{2}\hspace{0.1cm}(\text{mod}\hspace{0.05cm}e)$. Hence, when $0<m<2$, Eq.~(\ref{eq:1DTI}) is said to be in its obstructed atomic limit phase\cite{bradlyn2017topological}, whereas when $m<0$ or $m>2$, it is said to be in its trivial phase.

 To construct a single layer of our full three-dimensional insulator, we stack several copies of (\ref{eq:1DTI}) along the $y$-direction, and introduce a coupling strength $t_{y}$ along the stacking direction. This leads to the following Bloch Hamiltonian for each layer:
\begin{equation}\begin{split}\label{eq:2DWTI}
H_{\text{2DWTI}}(k_{x},k_{y})=H_{\text{1DTI}}(k_{x})-t_{y}\cos(k_{y})\sigma_{z}.
\end{split}\end{equation}
For $0<m<2$, as long no solution exists for (i) $\cos(k_{y})=-\frac{m}{t_{y}}$ and (ii) $\cos(k_{y})=\frac{2-m}{t_{y}}$, the system is a weak topological insulator. If for any value of $(m, k_x, k_y)$, a solution exists to either (i) \textit{or} (ii), the system becomes a Dirac semimetal with a single pair of Dirac nodes in the BZ (DSM 1 in Fig.~\ref{fig:2DWTIphasediagram}). Similarly, if solutions exist for both (i) \textit{and} (ii), then the system is a Dirac semimetal with two pairs of Dirac nodes in the BZ  (DSM 2 in Fig.~\ref{fig:2DWTIphasediagram}). These results can be summarized in a phase diagram for (\ref{eq:2DWTI}) shown in Fig.~\ref{fig:2DWTIphasediagram}. In both Dirac semimetal phases, the nodes are stabilized by a composite $TI$ symmetry (where $T=K$ is spinless time-reversal symmetry). For simplicity, we will consider the regime where $|m|<|t_{y}|$ \textit{or} $|2-m|<|t_{y}|$ so that each layer is in the DSM 1 phase. 

\subsection{Model Hamiltonian and Stacking Fault Procedure}
\label{subsec:modelhamiltonian}
\begin{figure}[ht]
\includegraphics[scale=0.22]{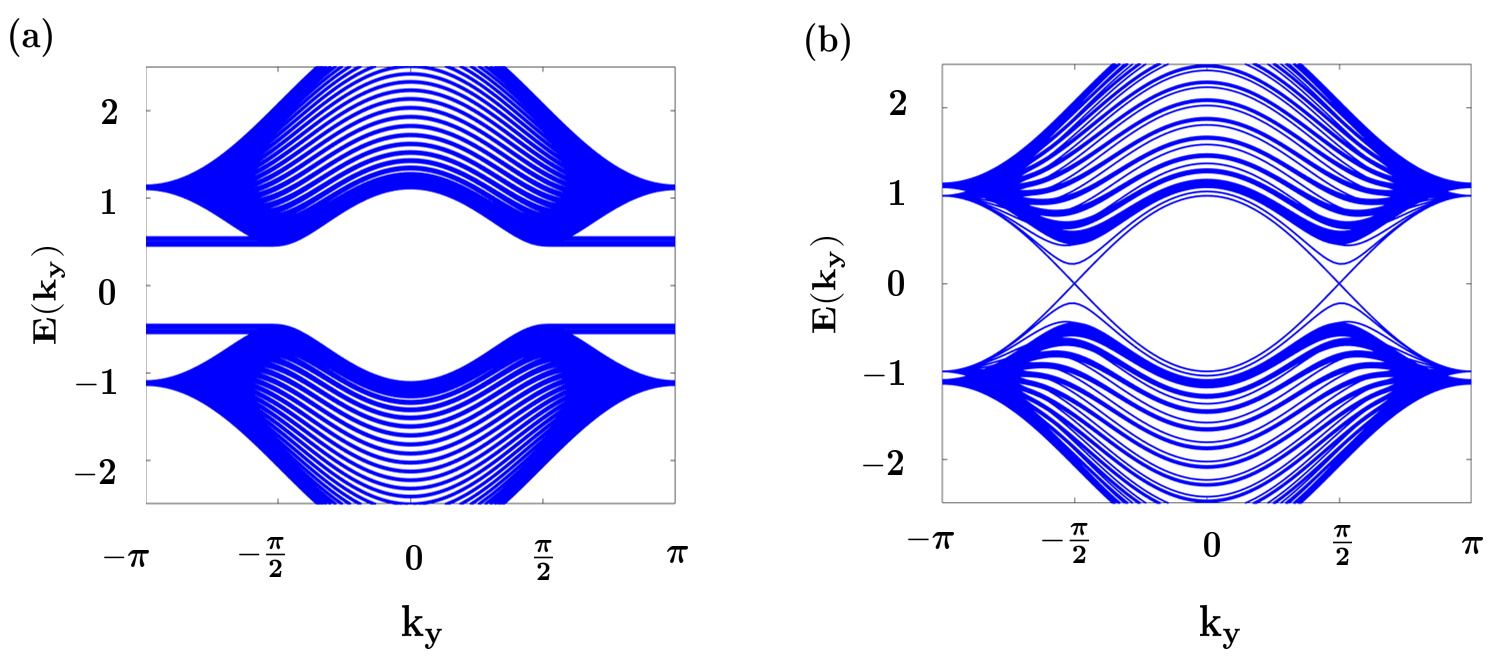}
\caption{Band structure of the pristine 3D bulk insulator given by Eqns.~(\ref{eq:3Dpristine1}) and (\ref{eq:3Dpristine2}) shown in (a), for $m=2$, $t_{y}=1$ with  $\gamma_{z}=0.05$ and $\delta_{z}=0.5$ with open boundaries in the $x$ direction and periodic boundaries in the $z$ direction, for $N_{z}=14$ unit cells ($28$ layers in total) and $N_{x}=28$ sites. The bulk band structure after the stacking fault procedure is performed is shown in (b) (periodic boundaries in both the $x$ and $z$ directions). The layer in the $7^{\text{th}}$ unit cell ($14^{\text{th}}$ layer) was extracted and the exposed layers are rejoined with coupling strength $\delta_{z}=0.5$. Dirac nodes appear in the bulk at $k_{y}=\pm\frac{\pi}{2}$.}
\label{fig:embeddedtdsmenergyspectrum}
\end{figure}

\indent To construct our 3D bulk insulator, we first layer copies of (\ref{eq:2DWTI}) along the $z$-direction such that each unit cell in the $z$-direction consists of two layers of (\ref{eq:2DWTI}). We denote these two layers as $A$ and $B$ which have model parameters $(m_{A},t_{y,A})$ and $(m_{B},t_{y,B})$ respectively. In this limit when the layers are decoupled the Bloch Hamiltonian for the pristine 3D system is
\begin{equation}\begin{split}\label{eq:3Dpristine1}
&H(\mathbf{k})=\sin(k_{x})(\mathbf{1}\otimes\sigma_{y})\\
&+(1-m-\cos(k_{x})-t_{y}\cos(k_{y}))(\mathbf{1}\otimes\sigma_{z}),
\end{split}\end{equation}
where the Pauli matrices $\tau_{i}$ ($i=x,y,z$) represent the layers within each unit cell. The Bloch Hamiltonian in (\ref{eq:3Dpristine1}) maintains the spinless time-reversal symmetry present in (\ref{eq:2DWTI}), as well as the inversion symmetry within the $xy$ plane, i.e., $C_{2z}$ symmetry represented by the operator $C_{2z}=\mathbf{1}\otimes\sigma_{z}$. When the model parameters $(m,t_{y})$ are tuned to the DSM 1 phase as stated in Subsec.~\ref{subsec:layerhamiltonian}, each layer is a Dirac semimetal with a pair of Dirac nodes in the bulk BZ. In order to gap out these Dirac nodes and turn (\ref{eq:3Dpristine1}) into an insulator, we introduce a coupling term  $H_{\text{coupling}}(\mathbf{k})$ that couples the layers of (\ref{eq:3Dpristine1}): 
\begin{equation}\begin{split}\label{eq:3Dpristine2}
&H_{\text{coupling}}(\mathbf{k})=(\gamma_{z}+\delta_{z}\cos(k_{z}))(\tau_{y}\otimes\sigma_{x})\\
&+(\delta_{z}\sin(k_{z}))(\tau_{x}\otimes\sigma_{x}),
\end{split}\end{equation}
where $\gamma_{z}$ and $\delta_{z}$ are the intra-cell and inter-cell couplings respectively. We note that in the presence of the coupling term the model has broken $T$ and $C_{2z},$ but the composite $C_{2z}T$ symmetry is maintained. Hence, when each individual layer of our 3D model is tuned to the DSM 1 phase in (\ref{eq:3Dpristine1}), the coupling Eq. (\ref{eq:3Dpristine2}) gaps the entire 3D system to form a trivial insulator. The open boundary spectrum of the 3D model with the coupling term included is illustrated in Fig.~\ref{fig:embeddedtdsmenergyspectrum} (a). The bulk, along with the typical Dirac semimetal edge states, are completely gapped out, indicating that the system is a trivial insulator in the absence of a stacking fault.

\begin{figure}[ht]
\includegraphics[scale=0.25]{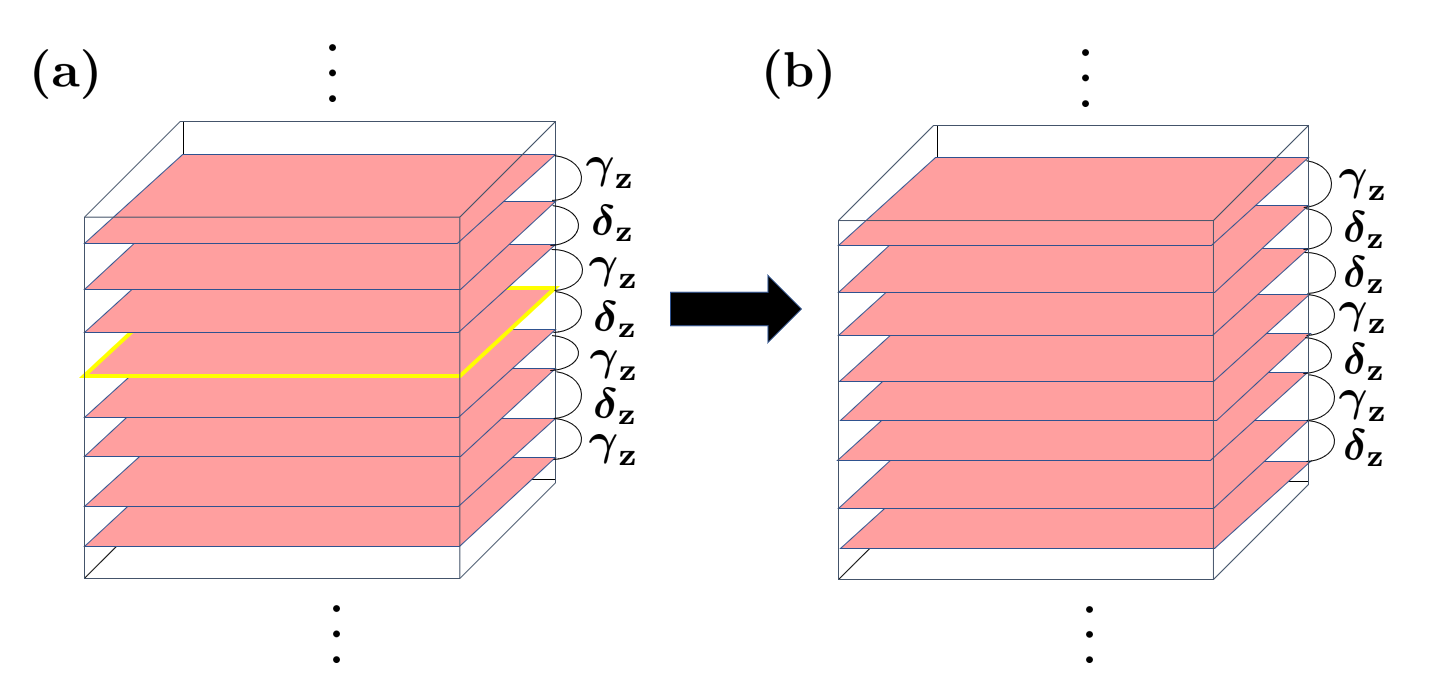}
\caption{Illustration of stacking fault procedure (Volterra construction) in the layered topologically trivial 3D insulator. In (a), the pristine 3D insulator is shown. The $B$ layer to be extracted is indicated by the yellow outline. Once this stacking fault is introduced, the resulting structure is shown in (b).}
\label{fig:stackingfaultpicture}
\end{figure}

 We now introduce a stacking fault into this system by extracting a single layer from the $\left(\frac{N_{z}}{2}\right)^{\text{th}}$ unit cell (i.e., the $N_{z}^{\text{th}}$ layer) (Fig.~\ref{fig:stackingfaultpicture}). The two exposed layers are then coupled with a coupling strength $\delta_{z}$, and the resulting band structure is shown in Fig.~\ref{fig:embeddedtdsmenergyspectrum}(b), which features Dirac nodes at $k_{y}=\pm\frac{\pi}{2}$. Therefore, by introducing a stacking fault, we have explicitly broken the unit cell structure of the pristine trivial insulator and generated an embedded Dirac semimetal. As we will see below it is also possible for other types of metallic features to emerge when partial dislocation defects are introduced into this trivial insulator (see Subsec.~\ref{subsec:partialdislocation}), but first we discuss the stability of the Dirac nodes in the ETSM.
 
\subsection{Characterizing the Topology of the ETSM}
\label{subsec:topology}
\indent In this subsection, we will demonstrate the topological nature of the embedded Dirac semimetal by calculating the Zak-Berry phase for a system with and without a stacking fault. Since we will consider the presence of a defect, we treat the $z$-direction in real-space and have an effective 2D BZ spanned by $(k_x, k_y).$  For each fixed value of $k_y$ in this 2D BZ we can consider a quasi-1D system extended along the $k_x$ direction. Hence for each fixed $k_y$ we can consider the Zak-Berry phase along $k_x$ by integrating the adiabatic connection along $k_{x}$\cite{zak1989berry}:
\begin{equation}\begin{split}\label{eq:ZakBerryphase}
\theta_{B}(k_{y})=\int\limits_{-\pi}^{\pi}dk_{x}\hspace{0.1cm}\text{tr}\hspace{0.05cm}\mathcal{A}_{x}(k_{x}),
\end{split}\end{equation}
where
\begin{equation}\begin{split}\label{eq:ZakBerryconnection}
\mathcal{A}_{x}(k_{x})=\langle u_{m}(k_{x},k_{y})|i\nabla_{k_{x}}|u_{n}(k_{x},k_{y})\rangle,
\end{split}\end{equation}
is the non-Abelian adiabatic connection derived from the occupied states (i.e., states with energy less than the Fermi-level $E_{F}=0$). For each value of $k_y$ this quasi-1D Berry phase is quantized by the $C_{2z}T$ symmetry and can take values of only $0$ or $\pi$ if $k_y$ does not pass through a gapless Dirac point (which we avoid in our plots).  We compute $\theta_{B}$ along a discretized grid of $k_{x}$ points at each $k_{y}$ value (excluding the values where the Dirac nodes occur at $k_{y}=\pm\frac{\pi}{2}$). To do this, we use the discretized gauge invariant Wilson loop expression
\begin{equation}\begin{split}\label{eq:discretizedZakBerryphase}
&\theta_{B}(k_{y})=\text{Im}\hspace{0.05cm}\text{log}\prod\limits_{i=1}^{N_{\text{grid},x}}\text{det}\hspace{0.1cm}\mathcal{U}_{i}(k_{y})\\
&[\mathcal{U}_{i}(k_{y})]_{mn}=\langle u_{m}(k_{x,i},k_{y})|u_{n}(k_{x,i+1},k_{y})\rangle.
\end{split}\end{equation}\\
\begin{figure}[ht]
\includegraphics[scale=0.26]{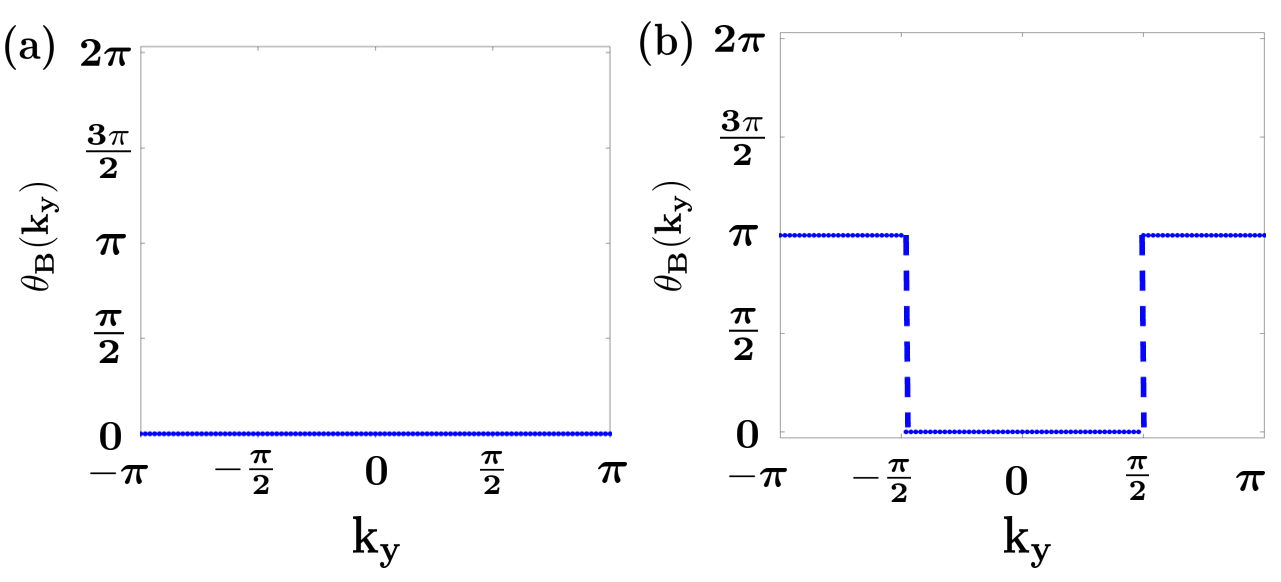}
\caption{Plot of the discretized (Zak)-Berry phase given by Eq.~(\ref{eq:discretizedZakBerryphase}) as a function of $k_{y}$, for a discretized 2D BZ of $N_{\text{grid},x}=100$ points along $k_{x}$ by $N_{\text{grid},y}=500$ points along $k_{y}$. Identical model parameters were used as in Fig.~\ref{fig:embeddedtdsmenergyspectrum}. The plot in (a) is when the system is a trivial insulator with no stacking fault and (b) corresponds to the system when the stacking fault is introduced. The dashed vertical lines indicate discontinuities where the Berry phase cannot be computed, which is precisely at the location of the Dirac nodes, $k_{y}=\pm\frac{\pi}{2}$.}
\label{fig:embeddedtdsmberryphase}
\end{figure}\\

A plot of the calculated Berry phase is shown in Fig.~\ref{fig:embeddedtdsmberryphase} as $k_{y}$ is tuned from $-\pi$ to $\pi.$ For Fig.~\ref{fig:embeddedtdsmberryphase} (a) we consider a system without a stacking fault and the Berry phase is constant across the $k_y$ BZ. For Fig.~\ref{fig:embeddedtdsmberryphase} (b) we show the results for a single stacking fault and find that the Berry phase remains quantized, but it changes its value by a quantized amount precisely at the locations of the Dirac nodes, $k_{y}=\pm\frac{\pi}{2}$. This result can be understood as follows. In computing the Berry phase across the BZ, we effectively treated the BZ as a family of 1D insulators parametrized by $k_{y}$. That is, each value of $k_{y}$ (except at $k_{y}=\pm\frac{\pi}{2}$) represents a 1D insulator oriented along the $x$-direction having an effective $y\to -y$ inversion symmetry generated by $C_{2z}T$. So as $k_{y}$ is tuned within $\left(-\frac{\pi}{2},\frac{\pi}{2}\right)$, the system is effectively in a trivial insulator phase since each 1D insulator has Berry phase $\theta_{B}=0$.  Likewise, the system is in a topological insulator (obstructed atomic limit) phase when $k_{y}$ is varied within $\left(-\pi,-\frac{\pi}{2}\right)$ and $\left(\frac{\pi}{2},\pi\right]$ since each 1D insulator has Berry phase $\theta_{B}=\pi$. Therefore, each Dirac node appearing in momentum space represents a transition between a trivial phase and a topological phase based on the Berry phase characteristics, making this an ETDSM. The key quantized Berry phase jump that illustrates the topological nature of the 2D Dirac cones is protected by the $C_{2z}T$ symmetry, and this symmetry also provides the local momentum-space stability of the defect-localized Dirac node degeneracy. We also characterize the topological nature of the embedded Dirac semimetal using the single-particle entanglement spectrum and by computing the entanglement Berry phase in analogy to Ref. \onlinecite{tuegel2018embedded}, which is discussed in Appendix B.

\subsection{T\lowercase{opological} M\lowercase{etallic} S\lowercase{tates} I\lowercase{nduced} \lowercase{by} \lowercase{a} P\lowercase{artial} D\lowercase{islocation}}
\label{subsec:partialdislocation}
\begin{figure}[ht]
\includegraphics[scale=0.3]{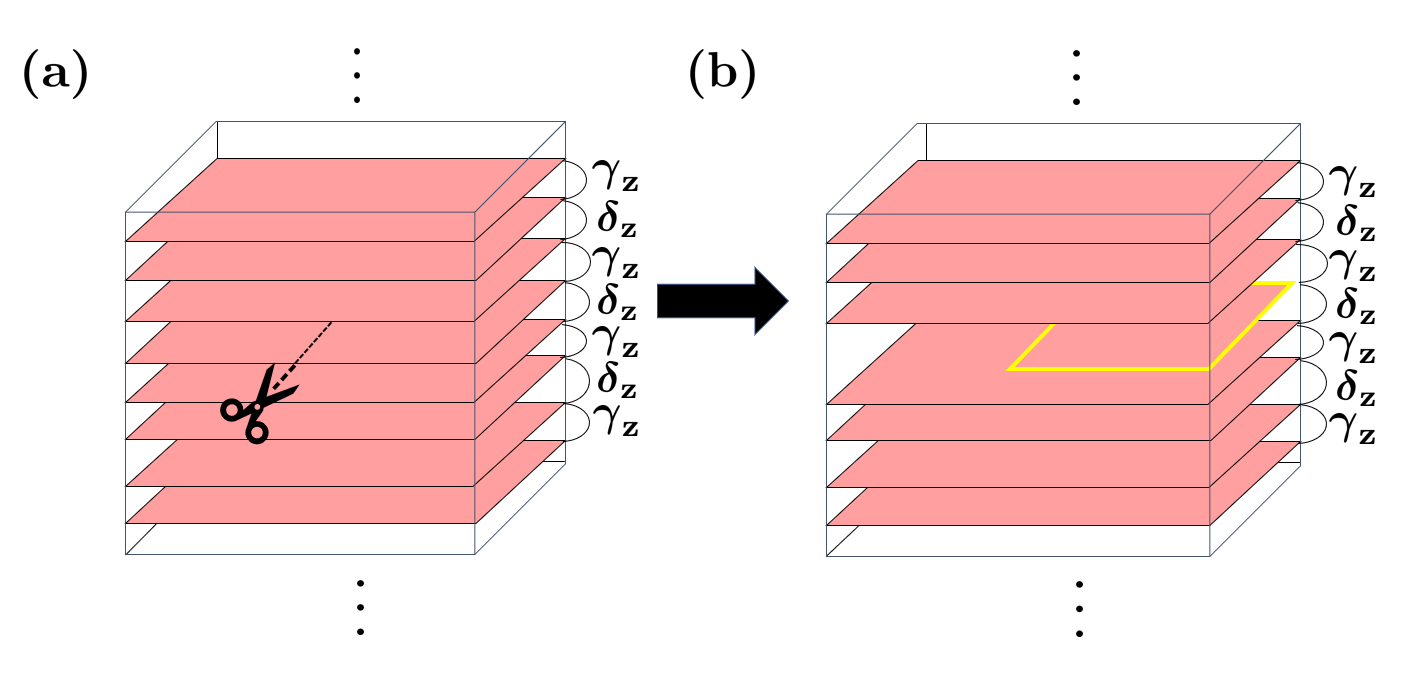}
\caption{Illustration of the partial dislocation procedure. (a) and (b) show how a partial dislocation is introduced in the pristine 3D topologically trivial insulator. A layer is cut in half parallel to the $y$-direction so that $k_{y}$ remains a good quantum number, and the layers above and below are recoupled. The band structure of the trivial 3D insulator before the partial dislocation is illustrated in (a) for periodic boundaries along the $z$ direction and open boundaries along the $x$ direction. As expected, the bulk is gapped and the edge states are gapped out. The bulk band structure after the partial dislocation is performed is shown in (b).}
\label{fig:partialdislocation}
\end{figure}

\begin{figure}[ht]
\includegraphics[scale=0.265]{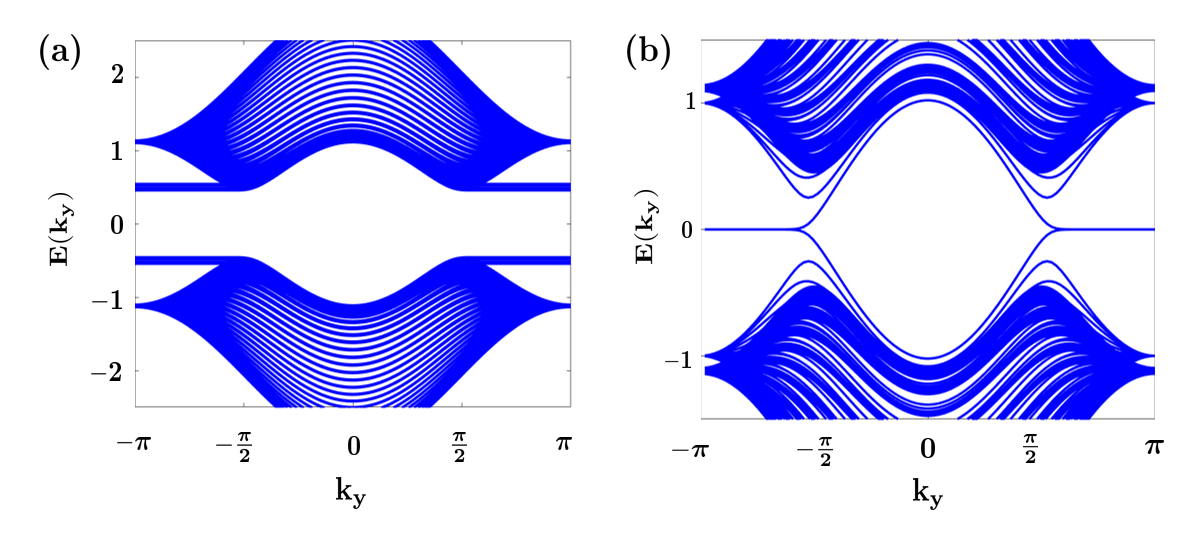}
\caption{Illustration of how topologically protected gapless modes can emerge from an ordinary insulator. As shown in (a) before the partial dislocation is created, the bulk is gapped and the edge states are gapped out. The bulk band structure after the partial dislocation is performed is shown in (b) and was plotted for periodic boundaries in both the $z$ and $x$ directions. Gapless modes intervene between $k_{y}=-\frac{\pi}{2}$ and $k_{y}=\frac{\pi}{2}$ that traverse the boundary of the BZ. Identical model parameters were used to construct the plots as in Fig.~\ref{fig:embeddedtdsmenergyspectrum}.}
\label{fig:partialdislocationbandstructure}
\end{figure}

In addition to stacking faults, partial dislocations (which can be heuristically thought of as terminated stacking faults) provide another example of a crystalline defect that can give rise to metallic states when introduced to a trivial insulating system. In particular, we show that metallic states that are topologically protected can emerge in a trivial background when a partial dislocation is applied to our 3D model. To implement the partial dislocation, we remove only half of the layer located in the $\left(\frac{N_{z}}{2}\right)^{\text{th}}$ unit cell (i.e., the $N_{z}^{\text{th}}$ layer) and the layers above and below are recoupled. Specifically, we perform the cut parallel to the $y$-direction so that translational invariance is preserved in this direction and $k_{y}$ remains a good quantum number. This procedure is illustrated in Fig.~\ref{fig:partialdislocation} (a) and (b), and the resulting energy spectrum is shown in Fig.~\ref{fig:partialdislocationbandstructure} (b) which can be compared with the defectless spectrum shown in Fig.~\ref{fig:partialdislocationbandstructure} (a). In the system with a partial dislocation we find a set of gapless modes  that connect between the embedded Dirac nodes at $k_{y}=\pm\frac{\pi}{2}$ and are located in position space on edge of the cleaved layer, i.e., on the partial dislocation line. These modes emerge in the system since the layer itself is a Dirac semimetal and these are the flatband edge states that connect the Dirac nodes. Hence, we have found that both stacking faults and partial dislocations give rise to embedded metallic states that emerge in the bulk trivial insulator.
\section{L\lowercase{andau} L\lowercase{evel} S\lowercase{pectrum} \lowercase{of} E\lowercase{mbedded} T\lowercase{opological} D\lowercase{irac} S\lowercase{emimetal}}
\label{sec:landaulevel}
\begin{figure}[ht]
\includegraphics[scale=0.45]{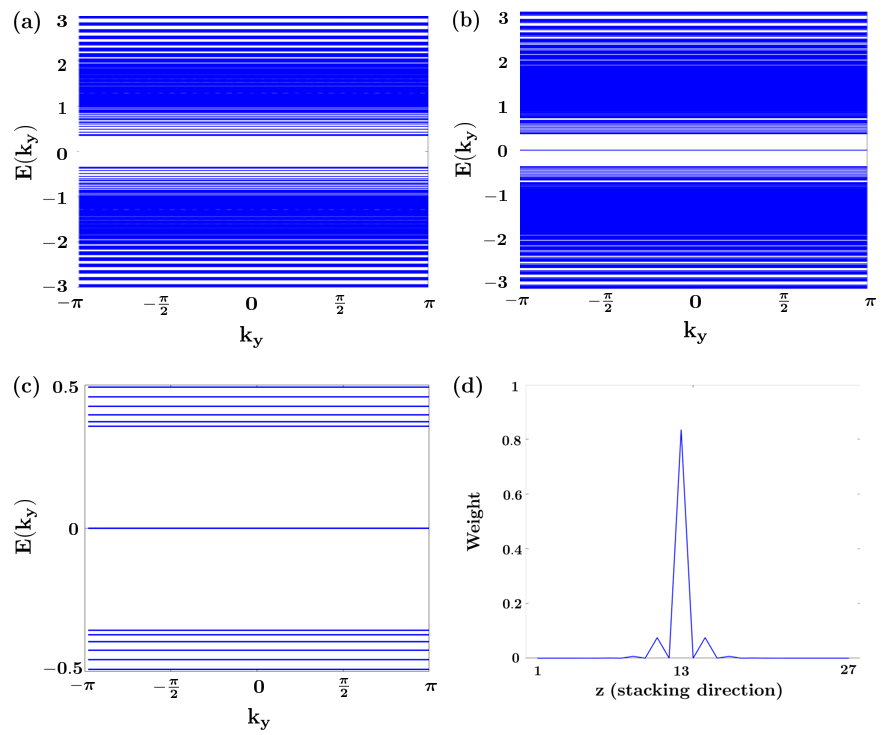}
\caption{Landau Level spectrum of (a) the trivial insulator before the stacking fault is introduced and (b) the ETDSM when the stacking fault is introduced. For each plot, $N_{x}=62$ sites and $q=31$ as required by magnetic translational invariance, with $m=2$, $t_{y}=1$, $\gamma_{z}=0.5$, and $\delta_{z}=0.15$ with $N_{z}=28$ layers. The Dirac LL spectrum is realized at this value of $q$ (e.g., low values of $B$) as shown in (b). The LL spacing changes from the spacing found in relativistic LL to the uniform spacing found in non-relativistic LL. Furthermore, a four-fold degenerate set of zero energy bands appears in the ETDSM spectrum of (b), and (c) is a zoomed-in plot of the relativistic LL in (b) illustrating the zero energy bands. (d) is a plot of the zero mode wavefunction probability distribution along the stacking direction. The zero modes are localized along the layer with the ETDSM after the stacking fault is introduced, as shown in (d).}
\label{fig:landaulevelspectrum}
\end{figure}

We now shift our focus to discussing a viable method for identifying an ETDSM in an experimental setting. In particular, we propose that an ETDSM can be readily identified by resolving its relativistic massless Dirac Landau level (LL) spectrum and the associated integer QHE sequence. In order to resolve the LL spectrum, the system with the stacking fault must be placed in a magnetic field with a non-vanishing component normal to the plane of the stacking fault. For simplicity we will consider a magnetic field along the $z$-direction, and use a Landau gauge $\mathbf{A}=Bx\hat{y}.$ We take the magnetic field strength to be $B=\phi\frac{\Phi_{0}}{a^{2}},$ where $\Phi_{0}=h/e$ is the flux quantum, and the lattice constant is denoted by $a$ along the $x$ and $y$ directions (we set both $\Phi_{0}$ and $a$ to unity). We let $\phi=\frac{p}{q}$ be a rational number\cite{Hofstadter1976} where $p$ gives the number of integer flux quanta per magnetic unit cell of size $q$. The technical details of the stacking fault Hamiltonian in a magnetic field are included in Appendix C.\\
\indent To correctly obtain the Dirac LL spectrum from a lattice model we must choose a large enough $q$ to recover the continuum result\cite{Bernevig2006a,Bernevig2007}. Diagonalizing this Hamiltonian for $\phi=1/31$ produces the Landau levels shown in Fig.~\ref{fig:landaulevelspectrum}. Fig.~\ref{fig:landaulevelspectrum} (a) shows the LL spectrum for the insulator given by Eqns.~(\ref{eq:3Dpristine1}) and (\ref{eq:3Dpristine2}) with no stacking fault, while Fig.~\ref{fig:landaulevelspectrum} (b) shows the LL spectrum for the ETDSM when the stacking fault is introduced. We see a clear zeroth LL (formed by four degenerate bands at $E=0$) and non-uniform LL spacing in Fig.~\ref{fig:landaulevelspectrum} (b) which is expected in a system with 2D Dirac points. These zero-energy bands are a key feature that distinguishes the trivial insulator from the ETDSM and signifies the presence of bulk Dirac nodes. To summarize, we find that when the ETDSM is subject to a magnetic field, the Dirac LL spectrum can be successfully resolved at low energies. Consequently, given the consistency of our numerical results presented in this section with those found in Refs.~\onlinecite{Bernevig2006a,Bernevig2007}, for a graphene lattice model, we expect the ETDSM to exhibit an associated relativistic integer QHE sequence, which could potentially be an experimentally accessible signature. We note that a similar proposal of measuring the unconventional QHE sequence was made in Ref.~\onlinecite{Slager2016} to identify time-reversal symmetry protected semimetals along grain boundaries in TBIs.
\section{C\lowercase{onclusion}}
\label{sec:conclusion}
In conclusion, we have investigated the viability of realizing new, potentially topological, (semi)-metallic phases in systems via defect insertions in ordinary insulators. We established the notion of ETSMs as topological semimetals localized on defects embedded in trivial insulating environments, by considering topologically trivial insulating systems in $D$ dimensions composed of coupled multi-layers $d$-dimensional topological semimetals. Starting with the simplest example in $D=2$ dimensions, we considered an ordinary insulator composed of layered $d=1$ spinless electron chains, and introduced two stacking faults, which unveiled an embedded metal with metallic Fermi points in its band structure. Extending this to $D=3$ dimensions, we considered an insulator consisting of coupled layers of 2D Dirac semimetals. Introducing a  stacking fault in this system resulted in an ETDSM, with visible Dirac nodes in the bulk band structure. We also showed the partial dislocations can bind the expected edge states of a 2D Dirac semimetal. Finally, we discussed how an ETSM can be identified in experiment by resolving the relativistic massless Dirac LL spectrum at low energies.\\
\indent Although embedded Weyl and line-node semimetals were not explicitly considered in this work, we note that the construction laid out here for embedded Dirac semimetals applies to these systems straightforwardly for $D=4$ dimensions. For example, an embedded Weyl semimetal can be realized in an topologically trivial insulator in $D=4$ dimensions composed of 3D layers, which themselves are coupled multi-layers of 2D ($d=2$) Chern insulators, by applying a stacking fault. However, because these systems require higher dimensionality beyond $D=3$ dimensions, their physical relevance will be limited to engineered materials having additional synthetic dimensions.\\
\\
\indent Finally, it warrants further examination whether embedded topological phases can be related to other types of higher-order topological phases. As mentioned in the introduction, it has been shown previously in\cite{Khalaf2018a,Matsugatani2018,Trifunovic2019} that HOTIs can be smoothly deformed to ETIs. However, it may be possible that higher-order topological semimetals (HOTSMs)\cite{Lin2018,Wang2020,Ghorashi2020,SImon2021} can also harbor ETIs. For example, the 3D higher-order Dirac semimetal discussed in\cite{Lin2018} and the higher-order Weyl semimetal discussed in \cite{Ghorashi2020} relies on coupled-layer constructions of 2D quadrupole models. If one considers a dimerized coupling structure along the stacking direction of the quadrupole models that breaks the translational symmetry but preserves crystalline symmetries such as $C_{4}$ and/or mirror symmetries, this can result in an embedded quadrupole insulator. For the ETSMs considered in this work, it may be viable to employ a similar type of argument to relate ETSMs to HOTIs or HOTSMs, which we leave to future study.
 
\section{A\lowercase{cknowledgments}}
\vspace{-0.29cm}
\noindent We thank Penghao Zhu and Ze-Min Huang for insightful discussions, and Nick Abboud for some useful initial discussions on this work. S.V. is supported by the NSF Graduate Research Fellowship Program under Grant No. DGE - 1746047. T.L.H. thanks ARO under the MURI Grant W911NF2020166.

\bibliographystyle{apsrev4-1}
\bibliography{EmbeddedBibliography}

\begin{appendix}

\section{B\lowercase{and} D\lowercase{ispersion} \lowercase{near} F\lowercase{ermi} L\lowercase{evel} \lowercase{vs}. S\lowercase{tacking} F\lowercase{ault} C\lowercase{oupling} S\lowercase{trength} f\lowercase{or} E\lowercase{mbedded} M\lowercase{etal} \lowercase{in} (1+1)-D}
\begin{figure}[ht]
\centering
\includegraphics[scale=0.37]{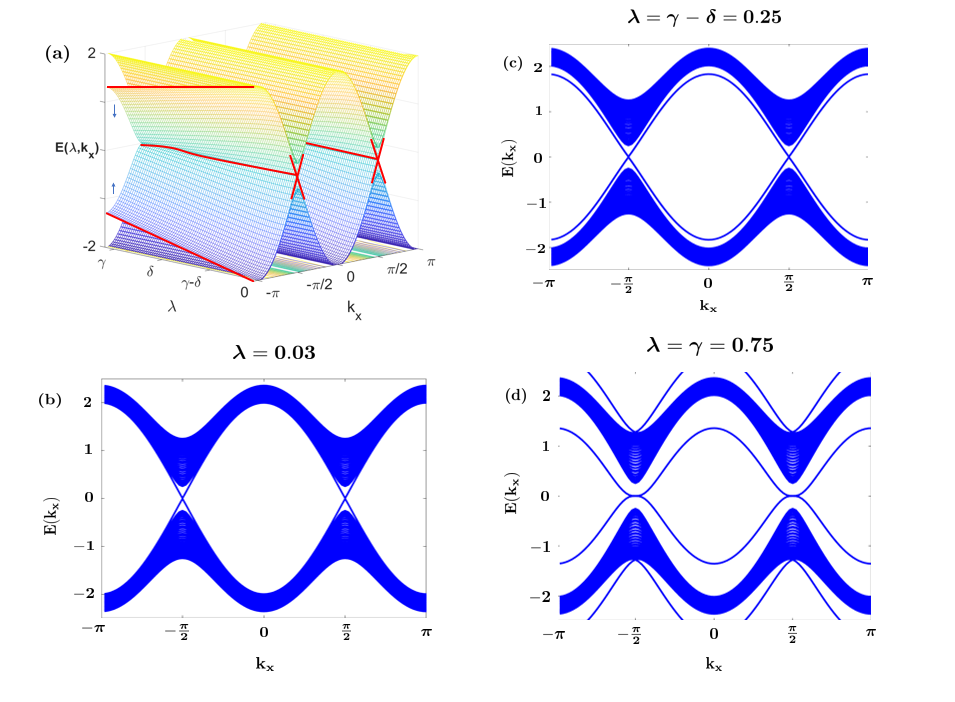}
\caption{Plots of the band dispersion near the Fermi level ($E_{F}=0$) versus the stacking fault coupling strength $\lambda$. We take $\gamma>\delta$ and use the same parameter values $\gamma=0.75$ and $\delta=0.5$ as in Fig.~\ref{fig:embeddedmetalenergyspectrum}. (a) Plot of the band dispersion near the Fermi level as a function of both $\lambda$ and $k_{x}$. The bolded red lines highlight how the bulk band edges bend towards each other as $\lambda$ is increased. (b)-(d) Plots of the band structure for various values of $\lambda$. When $\lambda$ is small, the dispersion of the nodes at $k_{x}=\pm\frac{\pi}{2}$ is linear. In particular, when $\lambda=0$, numerical calculation of the slope shows that $\frac{\Delta E}{t\Delta k_{x}}\approx 2.0000636$ (with lattice constant $a=1$) and for $\lambda=0.03$ (shown in (b)), $\frac{\Delta E}{t\Delta k_{x}}\approx 1.99366$. The dispersion of the nodes remains linear in (c), but the slope of the dispersion begins to shift noticeably. As $\lambda$ approaches the value of $\gamma>\delta$, the dispersion of the nodes is flattened. The intermediate value of $\lambda=\delta$ is illustrated in Fig.~\ref{fig:embeddedmetalenergyspectrum} b) in the main text.}
\label{fig:dispersionvsstackingfault}
\end{figure}

In this section, we present additional figures of the band dispersion near the Fermi level as a function of the stacking fault coupling strength. When the stacking fault procedure is performed, the exposed chains are taken to have generic coupling strength $\lambda$ where $0\leq\lambda\leq\max(\gamma,\delta)$. Figs.~\ref{fig:dispersionvsstackingfault} (a)-(d) provide a full illustration of the evolution of the band dispersion as $\lambda$ is tuned. As $\lambda$ is increased, the edges of the conduction and valence bands giving rise to the nodes at $k_{x}=\pm\frac{\pi}{2}$ approach each other. This is also accompanied by the fact that the dispersion of the nodes changes as $\lambda$ approaches $\max(\gamma,\delta)$. Namely, the slope of the dispersion near the Fermi level (i.e., the Fermi velocity) decreases as $\lambda$ increases.
\\
\indent We can gain some insight into the relationship between the band dispersion near the Fermi level at $k_{x}=\pm\frac{\pi}{2}$ and the stacking fault coupling strength $\lambda$ by considering the Hamiltonian given by Eq.~(\ref{eq:pristineBlochhamiltonian2D}) in one of two ``dimerized" limits: (i) $\gamma\neq 0$, $\delta=0$ and (ii) $\gamma=0$, $\delta\neq 0$. Below, we consider the dimerized limit given by (i), but the results are exactly the same for (ii) by simply replacing $\gamma$ with $\delta$. Performing a $k\cdot p$ expansion, the dispersion near $k_{x}=\pm\frac{\pi}{2}$ has the following form,
\begin{equation}\begin{split}\label{eq:kdotpdispersion}
E_{\pm}(\delta k_{x})\approx\pm\left(\alpha(\gamma,\lambda)\delta k_{x}+\beta(\gamma,\lambda)\delta k_{x}^{3}\right)
\end{split}\end{equation}
\noindent where $\delta k_{x}=k_{x}\pm\frac{\pi}{2}$ is the deviation in momentum from the location of the nodes and the coefficients $\alpha(\gamma,\lambda)$ and $\beta(\gamma,\lambda)$ are given by,
\begin{equation}\begin{split}\label{eq:kdotpdispersioncoefficients}
&\alpha(\gamma,\lambda)=\frac{2t(\gamma^{2}-\lambda^{2})}{\gamma^{2}+\lambda^{2}}\\\\
&\beta(\gamma,\lambda)=-\frac{t(\gamma^{8}+2\gamma^{6}\lambda^{2}-192t^{2}\gamma^{2}\lambda^{4}-2\gamma^{2}\lambda^{6}-\lambda^{8})}{3(\gamma^{2}+\lambda^{2})^{4}}
\end{split}\end{equation}
\noindent When the stacking fault coupling is very weak (i.e., $\lambda\ll\gamma$), the coefficient $\alpha(\gamma,\lambda)$ is much greater than the coefficient $\beta(\gamma,\lambda)$ and the dispersion near $k_{x}$ is dominated by the first term in (\ref{eq:kdotpdispersioncoefficients}) and is linear. But as $\lambda$ is increased and specifically, when $\lambda$ becomes very close in value to $\gamma$, $\alpha(\gamma,\lambda)$ is significantly smaller than $\beta(\gamma,\lambda)$. Therefore, at strong stacking fault coupling $\lambda$ that is comparable to $\gamma$, the second term in (\ref{eq:kdotpdispersion}) dominates and leads to a cubic dispersion in energy near $k_{x}=\pm\frac{\pi}{2}$. In fact, at $\lambda=\gamma$, the band dispersion is purely cubic. This leads to the following relationship between the stacking fault coupling $\lambda$ and the magnitude of the group velocity near the Fermi level $v_{F}$,
\begin{equation}\begin{split}
v_{F}=\left|\frac{\delta E}{\delta k_{x}}\right|\approx\alpha(\gamma,\lambda)+\beta(\gamma,\lambda)\delta k_{x}^{2}
\end{split}\end{equation}
\noindent which shows that at weak $\lambda$, the group velocity is constant and at strong $\lambda$, the group velocity has a quadratic dispersion in momentum.

\section{T\lowercase{opological} C\lowercase{haracterization} \lowercase{of} ETSM \lowercase{from} E\lowercase{ntanglement} S\lowercase{pectrum}}
In this section, we provide an alternative means for identifying the topological characteristics of the embedded Dirac semimetal utilizing the single-particle entanglement spectrum. Using the same procedure as in Subsec.~\ref{subsec:topology}, we analyze entanglement Berry phase characteristics across the 2D BZ.
\subsection{Brief Review of Embedded Topological Insulators (ETIs) and Entanglement Spectrum (ES)}
\indent To motivate this section and understand why the entanglement spectrum is a useful tool for diagnosing ETSMs, we briefly review the concept of an Embedded Topological Insulator (ETI) from Ref.~\onlinecite{tuegel2018embedded}. Consider an ordinary (topologically trivial) environment $\mathcal{M}$ and a topological insulator (TI) $\mathcal{N}$ embedded in $\mathcal{M}$, and consider the case where $\text{dim}(\mathcal{M})>\text{dim}(\mathcal{N})$. Nominally, if a physical cut is made between the TI $\mathcal{N}$ and its trivial environment $\mathcal{M}$, topologically protected modes will emerge, such as surface states. However, if $\mathcal{N}$ is coupled to its environment $\mathcal{M}$, it becomes non-trivial to diagnose the topology of $\mathcal{N}$. At weak coupling it is possible to maintain these surface modes as long as $\mathcal{N}$ remains gapped and symmetries are preserved. However, at strong coupling $\mathcal{N}$ could potentially no longer be gapped, destabilizing the TI $\mathcal{N}$. In this latter case, the topology of $\mathcal{N}$ can be determined by disentangling $\mathcal{N}$ from its environment $\mathcal{M}$ through the use of the entanglement spectrum.\\
\indent For non-interacting fermionic systems harboring ETIs, the entanglement spectrum is derived as follows. First, a region $\mathcal{R}$ is identified as harboring the presumed ETI, and an entanglement cut is performed between $\mathcal{R}$ and its complementary region $\mathcal{R}^{c}$. This leads to a 1-body reduced density matrix $\rho_{\mathcal{R}}$ is determined by restricting the projector to the region $\mathcal{R}$ (i.e., $\rho_{\mathcal{R}}=P|_{\mathcal{R}}$). The entanglement spectrum (ES) is the eigenvalue spectrum $\{\lambda_{i}\}$ of the reduced density matrix $\rho_{\mathcal{R}}$ and is distributed over the unit interval $\{\lambda_{i}\}\subset[0,1]$. If the ES is gapped at $\lambda=\frac{1}{2}$, the set of eigenstates of $\rho_{\mathcal{R}}$ with $\lambda_{i}>\frac{1}{2}$ are identified as having the dominant contribution to correlations within $\mathcal{R}$ (i.e., ``occupied" entanglement states). The spectral projector is then constructed as
\begin{equation}\begin{split}\label{eq:entanglementspectralprojector}
\Pi_{\mathcal{R}}=\sum\limits_{\lambda_{i}>\frac{1}{2}}|\lambda_{i}\rangle\langle\lambda_{i}|.
\end{split}\end{equation}
\noindent If the topological index computed from $\Pi_{\mathcal{R}}$ is non-zero, then $\mathcal{R}$ is said to be topologically non-trivial. Furthermore, if any (symmetry-preserving) entanglement cuts made through $\mathcal{R}$ leads to a gapless entanglement spectrum, then $\mathcal{R}$ is said to harbor an ETI.
\subsection{Entanglement Berry Phase Analysis of the Embedded Dirac Semimetal}
\begin{figure}[ht]
\includegraphics[scale=0.75]{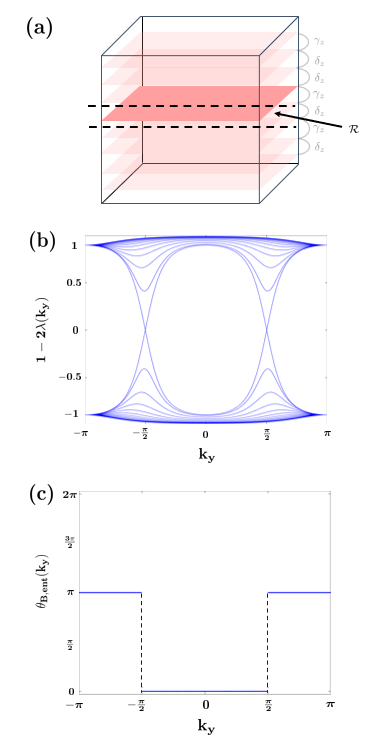}
\caption{Characterizing the topology of the embedded Dirac semimetal from the entanglement perspective. (a) illustrates the entanglement cuts and the region $\mathcal{R}$ chosen after the stacking fault procedure is performed (the layers above and below are faded out to emphasize the layer lying on the stacking fault after the extraction). In (b), the gapless entanglement `band structure' is shown. In (c), the entanglement Berry phase as a function of $k_{y}$ is shown, with the dashed lines indicating discontinuities in which the phase sharply changes value. These occur precisely at $k_{y}=\pm\frac{\pi}{2}$, precisely where the gapless points in the entanglement `band structure' occur. Identical model parameters were used to construct the plots in (b) and (c) as in the main text.}
\label{fig:embeddedtdsmentanglementspectrum}
\end{figure}

We now compute the entanglement spectrum of the embedded topological Dirac semimetal discussed in Sec.~\ref{sec:embeddeddiracsemimetal}. First, we take the spectral projector to be $P(k_{x},k_{y})=\sum\limits_{E_{n}(k_{x},k_{y})<E_{F}}|u_{n}(k_{x},k_{y})\rangle\langle u_{n}(k_{x},k_{y})|$ where the Fermi level is taken to be $E_{F}=0$. Before proceeding, we note that the band structure for the embedded Dirac semimetal is gapless precisely at $(k_{x},k_{y})=\left(\pm\pi,\pm\frac{\pi}{2}\right)$, which means our construction of the spectral projector at these points is not well-defined. Therefore, we modify our construction of the spectral projector by specifically avoiding the points where the band structure is gapless, i.e.
\begin{equation}\begin{split}\label{eq:spectralprojector}
P(k_{x},k_{y})=\sum\limits_{E_{n}(k_{x},k_{y})<E_{F}}|u_{n}(k_{x},k_{y})\rangle\langle u_{n}(k_{x},k_{y})|.
\end{split}\end{equation}
The sum is performed over all $(k_{x},k_{y})\neq\left(\pi,\pm\frac{\pi}{2}\right)$. With this definition of the spectral projector, we are ensuring that we consider only the eigenstates with energies strictly below the Fermi level and the calculation takes place in only the gapped regions of the band structure. This is crucial in evaluating the entanglement spectrum. From here, we construct the reduced density matrix $\rho_{\mathcal{R}}$ by restricting $P$ to the region $\mathcal{R}$ illustrated in Fig.~\ref{fig:embeddedtdsmentanglementspectrum} (a). The entanglement cuts are made directly above and below the layer where the stacking fault is located, which are above and below the $(N_{z}-1)^{\text{th}}$ layer specifically following the stacking fault procedure detailed in Subsec.~\ref{subsec:modelhamiltonian}. After computing the eigenvalue spectrum $\{\lambda_{i}\}$ of $P|_{\mathcal{R}}$, we construct the spectral projector over the entanglement eigenstates according to (\ref{eq:entanglementspectralprojector}). The resulting entanglement `band structure' constructed from the eigenvalues $\{\lambda_{i}\}$ is shown in Fig.~\ref{fig:embeddedtdsmentanglementspectrum} (b) and is clearly gapless, indicating that the region $\mathcal{R}$ we chose is harboring an ETSM.\\
\indent Just as was outlined in Subsec.~\ref{subsec:topology}, we divide the 2D BZ into 1D layers extended along the $k_{x}$ direction and treat $k_{y}$ as a layering parameter (we also note that a similar method for computing the entanglement spectrum in Weyl semimetals was also developed in Ref.~\onlinecite{Wang2014}). We evaluate the entanglement Berry phase and entanglement Berry connection, which we define as follows,
\begin{equation}\begin{split}\label{eq:entanglementZakBerryphase}
\theta_{B,\text{ent}}(k_{y})=\int\limits_{-\pi}^{\pi}dk_{x}\hspace{0.1cm}\text{tr}\hspace{0.05cm}\mathcal{A}_{x}(k_{x})
\end{split}\end{equation}
\begin{equation}\begin{split}\label{eq:entanglementZakBerryconnection}
\mathcal{A}_{x}(k_{x})=\langle \lambda_{m}(k_{x},k_{y})|i\nabla_{k_{x}}|\lambda_{n}(k_{x},k_{y})\rangle
\end{split}\end{equation}
When implementing (\ref{eq:entanglementZakBerryphase}) and (\ref{eq:entanglementZakBerryconnection}) in practice, we use the discretized gauge invariant Wilson loop expression over the entanglement eigenstates, which is generalized straightforwardly from Eq.~(\ref{eq:discretizedZakBerryphase}) in the main text. The result of this is illustrated in Fig.~\ref{fig:embeddedtdsmentanglementspectrum} (c), and the results are identical to Fig.~\ref{fig:embeddedtdsmberryphase} in the main text. As $k_{y}$ is varied from $(-\pi,\pi]$, the entanglement Berry phase is quantized throughout and only undergoes a change in value at $k_{y}=\pm\frac{\pi}{2}$.
\section{D\lowercase{etails} \lowercase{of} M\lowercase{odel} \lowercase{for} ETDSM \lowercase{in} \lowercase{a} M\lowercase{agnetic} F\lowercase{ield}}
\vspace{-0.3cm}
In this section, we provide the technical details for the ETDSM model when placed in a magnetic field. We introduce a magnetic field along the $z$-direction by working in the Landau gauge $\mathbf{A}=Bx\hat{y}$. The vector potential is incorporated into the Hamiltonian via the Peierls substitution,
\begin{equation}\begin{split}\label{eq:Peierlssubstitution}
&c_{n_{x}',n_{y}',n_{z}'}^{\dagger}c_{n_{x},n_{y},n_{z}}\\
&\to c_{n_{x}',n_{y}',n_{z}'}^{\dagger}c_{n_{x},n_{y},n_{z}}\exp\left\{i\int\limits_{(n_{x},n_{y},n_{z})}^{(n_{x}',n_{y}',n_{z}')}\mathbf{A}\cdot\hspace{0.05cm}d\mathbf{r}\right\}
\end{split}\end{equation} 
The magnetic field strength $B$ is taken to be $B=\phi\left(\frac{\Phi_{0}}{a^{2}}\right)$ where $p$ and $q$ are co-prime integers, $\Phi_{0}$ is the flux quantum, and $a$ is the lattice constant along the $x$ and $y$ directions (we take $\Phi_{0}=a=1$). The flux per plaquette in the $x-y$ plane is given by $\phi$, where $p$ prescribes the integer number of flux quanta per magnetic unit cell of size $q$ along the $x$-direction. Given the choice of Landau gauge, the standard translational invariance along the $x$-direction is no longer present. Instead, a magnetic translational invariance is present in which the Hamiltonian is invariant under translations of the form $n_{x}\to n_{x}+q$. Standard translational invariance is maintained along the $y$-direction under this choice of Landau gauge. Performing the Peierls substitution and taking a Fourier transform along the $y$ direction, the stacking fault Hamiltonian is now
\vspace{-0.1cm}
\begin{equation}\begin{split}\label{eq:stackingfaultHamiltonian}
H=H_{1}+H_{2}+H_{3}
\end{split}\end{equation}
where,
\begin{equation}\begin{split}\label{eq:stackingfaulthamiltonian21}
&H_{1}=\sum\limits_{n_{x}}\sum\limits_{k_{y}}\sum\limits_{n_{z}=1}^{N_{z}-1}\left(c_{n_{x},k_{y},n_{z}}^{\dagger}[H_{\text{on-site}}(n_{x},k_{y},\phi)]c_{n_{x},k_{y},n_{z}}\right.\\
&\left.\vphantom{\sum\limits_{n_{x}}\sum\limits_{k_{y}}}+c_{n_{x}+1,k_{y},n_{z}}^{\dagger}[H_{+\hat{x}}]c_{n_{x},k_{y},n_{z}}+\text{h.c.}\right)
\end{split}\end{equation}
\vspace{-0.1cm}
\begin{equation}\begin{split}\label{eq:stackingfaulthamiltonian22}
&H_{2}=\sum\limits_{n_{x}}\sum\limits_{k_{y}}\sum\limits_{n_{z}=1}^{\frac{N_{z}}{2}-1}\left(i\gamma_{z}c_{n_{x},k_{y},2n_{z}}^{\dagger}c_{n_{x},k_{y},2n_{z}-1}\right.\\
&\left.\vphantom{\sum\limits_{n_{x}}\sum\limits_{k_{y}}}+i\delta_{z}c_{n_{x},k_{y},2n_{z}+1}^{\dagger}c_{n_{x},k_{y},2n_{z}}+\text{h.c.}\right)
\end{split}\end{equation}
\vspace{-0.1cm}
\begin{equation}\begin{split}\label{eq:stackingfaulthamiltonian23}
&H_{3}=\sum\limits_{n_{x}}\sum\limits_{k_{y}}\sum\limits_{n_{z}=\frac{N_{z}}{2}}^{N_{z}-1}\left(i\gamma_{z}c_{n_{x},k_{y},2n_{z}+1}^{\dagger}c_{n_{x},k_{y},2n_{z}}\right.\\
&\left.\vphantom{\sum\limits_{n_{x}}\sum\limits_{k_{y}}}+i\delta_{z}c_{n_{x},k_{y},2n_{z}}^{\dagger}c_{n_{x},k_{y},2n_{z}-1}+\text{h.c.}\right)
\end{split}\end{equation}
where we have introduced periodic boundaries in the $x$-direction (i.e., $n_{x}+N_{x}\equiv n_{x}$) and open boundaries in the $z$ direction. In (\ref{eq:stackingfaulthamiltonian21}), we have introduced the following quantities,
\begin{equation}\begin{split}
[H_{\text{on-site}}(n_{x},k_{y},\phi)]=(1-m-t_{y}\cos(2\pi\phi n_{x}-k_{y}))\sigma_{z}
\end{split}\end{equation}
\begin{equation}\begin{split}
[H_{+\hat{x}}]=\frac{i\sigma_{y}-\sigma_{z}}{2}
\end{split}\end{equation}
Note that in (\ref{eq:stackingfaulthamiltonian21})-(\ref{eq:stackingfaulthamiltonian23}), in order for magnetic translational invariance to hold, we must have $N_{x}$ be divisible by the magnetic unit cell size $q$ (i.e., $N_{x}=lq$ for some $l\in\mathbb{Z}$). Furthermore, for numerical calculation purposes, we take the values of $k_{y}$ to be over a discrete grid with spacing $\frac{2\pi}{N_{y}}$ such that $k_{y}\in\left\{-\pi\left(1-\frac{2}{N_{y}}\right),\ldots,\pi\right\}$.

\end{appendix}

\end{document}